# Network Coding for Energy-Efficient Distributed Storage System in Wireless Sensor Networks

**Wang Lei [1], Yang Yuwang [1], Zhao Wei [1] and Lu Wei[1]**

[1] Computer Science School, Nanjing University of Science and Technology

Nanjing - China

[e-mail: yuwangyang@mail.njust.edu.cn]

*Corresponding author: Yang Yuwang

**Abstract:** A network coding-based scheme is proposed to improve the energy efficiency of distributed storage systems in WSNs (wireless sensor networks), which mainly focuses on two problems: firstly, consideration is given to effective distributed storage technology in WSNs; secondly, we address how to repair the data in failed storage nodes with less resource. For the first problem, we propose a method to obtain a sparse generator matrix to construct network codes, and this sparse generator matrix is proven to be the sparsest. Benefiting from the sparse generator matrix, the energy consumption required to implement distributed storage is reduced. For the second problem, we designed a network coding-based iterative repair method, which adequately utilizes the idea of re-encoding at intermediate nodes from network coding theory. Benefiting from the re-encoding, the energy consumption required by data repair is significantly reduced. Moreover, we provide an explicit lower bound of field size required by this scheme, which implies that this scheme can work over a very small field and the required computation overhead of coding is very low. The simulation result verifies that by using our scheme, the total energy consumption required to implement distributed storage system in WSNs can be reduced on the one hand, and on the other hand, this method can also balance energy consumption of the networks.

## 1 Introduction

How to improve the energy efficiency [1][2] is an important issue for network communication, especially in wireless multi-hop networks such as wireless sensor networks (WSNs) and ad hoc networks. This paper focuses on the energy efficiency of DSS (Distributed Storage System) in WSNs. In order to prevent permanent loss of data caused by failed storage nodes in WSNs, the distributed redundant storage technology [3] should be applied to the data perceived by the sensory nodes. In addition, when a storage node fails in the network, in order to maintain the fault-tolerance level of the storage system, the data in the failed storage node should be recovered at a new storage node (called newcomer). Alternatively, a new storage node equivalent to the failed node in function should be provided, so that the service level can be maintained. Energy efficiency in WSNs is mainly reflected on two aspects: total energy consumption of the network and the load balance of energy. Therefore, these two properties are the main metrics used to evaluate other existing methods as well as the method in this paper.

The most original distributed redundant storage technology is replication, and the Google file system [4] and Hadoop distributed file system [5] are its representative products. In a distributed storage system based on replication, in order to tolerate failure of one or more storage nodes, multiple copies of each data block must be generated. From the perspective of data recovery efficiency, this is definitely the most effective method. However, the defect of this method is high data redundancy, so the bandwidth and storage overhead required in this method is huge. Especially in WSNs with limited energy and resources, this defect becomes more serious.

Erasure coding [6][7] is another useful technology for distributed storage. Erasure coding transforms a message of $K$ symbols into a longer message with $N$ symbols such that the original message can be recovered from any $K$ out of the $N$ symbols. After using erasure coding in DSS, the original $K$ data files are encoded into $N$ sub-files, in which any $K$ ($K < N$) sub-files suffice to decode the original data. After encoding, the size of each sub-file equals the file size before the encoding, and this method can tolerate failures of up to ($N - K$) storage nodes. Moreover, the total size of the original $K$ data files is $M$, and the total size of the encoded sub-files is $NM / K$, so erasure coding can significantly reduce redundancy without sacrificing any reliability compared with replication. Due to this reason, erasure coding has been widely used in practical applications. The common RS (Reed-Solomon) code is a type of erasure code. However, using RS code may have the following two problems: first of all, this method used to generate sub-files through the generator matrix of RS code has a low energy efficiency, which will be elaborated on in Section 3; secondly, when one storage node fails, it is required to repair the failed data at a newcomer through participation of other surviving storage nodes. To this end, the new storage node needs to download at least $K$ different sub-files to recover the original files, select a regenerating codeword and regenerate the new sub-file through the original files and the regenerating codeword. Therefore, during the repair process, at least $K$ sub-files have to be downloaded, which requires a huge bandwidth overhead.

Therefore, how to improve these two problems without reducing the service level of the DSS becomes an interesting topic. Recently, the network coding technology proposed by Ahlswede et al. [8] has been believed to be a promising technology to address these two problems. The key idea of network coding is to re-encode at intermediate nodes. Based on this idea, the performance of multicast network (such as the throughput and energy efficiency) can be improved. Li et al.[9][9] proved that the network multicast capacity can be realized by using linear network coding, where the multicast capacity equals the smallest maximal flow from the source node to different sink nodes in value. After that, randomized linear network coding [10-12] and deterministic linear network coding [13][14] are proposed and applied in practical networks.

This paper uses network coding to improve the two problems mentioned above. The network coding technology generally includes two main aspects: encoding at the source node and re-encoding at intermediate nodes. Related theories and technologies of network coding theory can be used to improve the two problems: firstly, the theory of encoding at the source node corresponds to the distributed data storage technology, so the optimization theory of source node coding can be used to improve the performance of storage technology in WSNs, such as to reduce required transmissions, reduce total energy consumption of the network and balance energy consumption of individual nodes; secondly, when some storage nodes in the WSNs suffer from permanent failures, it is required to recover data of the failed node (or data with equivalent function) at newcomers through coordination with other surviving storage nodes, and the idea of re-encoding at intermediate nodes can significantly reduce the required bandwidth for repair, balance the network load and increase the security.

This paper has the following structure: in the second section, some related studies will be introduced, and the main contributions of this paper will be summarized; in the third and fourth sections, the storage and repair technologies based on network coding will be elaborated on, respectively; in the fifth section, we will provide an explicit proof to show that this scheme can work over a very small finite field; in the sixth section, the simulation and experimental results will be provided and discussed, and the last section will summarize the main conclusions.

## 2 Related studies

On the aspect of optimal storage technology, our method is mainly inspired by the work of Dimakis et al.[15][16][18]. In their study[15], they pointed out that the network codes built by the sparse generator matrix can help reduce the times of data transmissions required to realize distributed storage, so that the required energy to transmit packets can be reduced. In addition, when the sparse generator matrix is used for coding, it can also reduce the time complexity required by decoding. In this paper, we propose an effective method to construct the sparse generator matrix and prove that the constructed sparse matrix is the sparsest matrix with the MDS (Maximum Distance Separable) property, which means that the matrix cannot be sparser, because a sparser generator matrix will make it lose the MDS property. Therefore, theoretically speaking, when this matrix is used in the WSNs, the energy efficiency of the network will be optimal.

On the aspect of network-coding-based optimal repair technology, there is a significant difference between previous studies [15-19] and ours. In these previous studies, the researchers introduced network coding into data repair by regarding the repair problem as a single-source multicast problem in the network coding theory. Based on this idea, Dimakis et al. [18] discovered a tradeoff between the amount of storage in each node and the bandwidth required in the repair process, and MSR (Minimal Storage Repair) and MBR (Minimal Bandwidth Repair) are two special cases. This paper also proposes a network coding-based repair technology for DSS in WSNs, but our method is from a different perspective. The reason why we consider our method is related to network coding is that the core idea of re-encoding from network coding is employed at intermediate nodes for data repair. Benefiting from the re-encoding at intermediate nodes, the repair bandwidth is significantly reduced. In those previous studies, it is required that the newcomer must connect to multiple storage nodes. For example, in [17], each newcomer is required to connect to all the surviving storage nodes during the repair process. In this paper, there is no such requirement, and it is able to regenerating the data as long as the newcomer could connect to one storage node.

Recently, we designed and implemented a prototype system [20] based on network coding for the storage applications on the Internet. During the system experiments, we found that the system performance of DSS can be improved on many aspects by using the network coding technology, such as the bandwidth efficiency, load balancing and security. In spite of this, due to the architectural feature of Internet, we observed that the advantages of network coding cannot be adequately reflected. When the distributed storage technology based on network coding is used in wireless networks, the benefit of network coding can be adequately reflected due to the multi-hop and self-organization features of wireless networks. Therefore, this paper is a further study based on the development experience.

Our contributions can be summarized as follows: firstly, the energy efficiency of distributed storage system in WSNs is improved by obtaining a sparsest generator matrix; secondly, we propose an iterative repair method for data repair of DSS, which significantly reduces the energy consumption for repair; thirdly, we provide an explicit lower bound of finite field for the scheme in this paper, which suggests that the computation overhead of coding and decoding would be low.

## 3 Network coding-based distributed storage technology for WSNs

Previous researchers [15] pointed out that constructing a sparse generator matrix with the MDS property can help reduce the transmission times of network data, so that energy consumption can be accordingly reduced. This idea can be explained by the following examples.

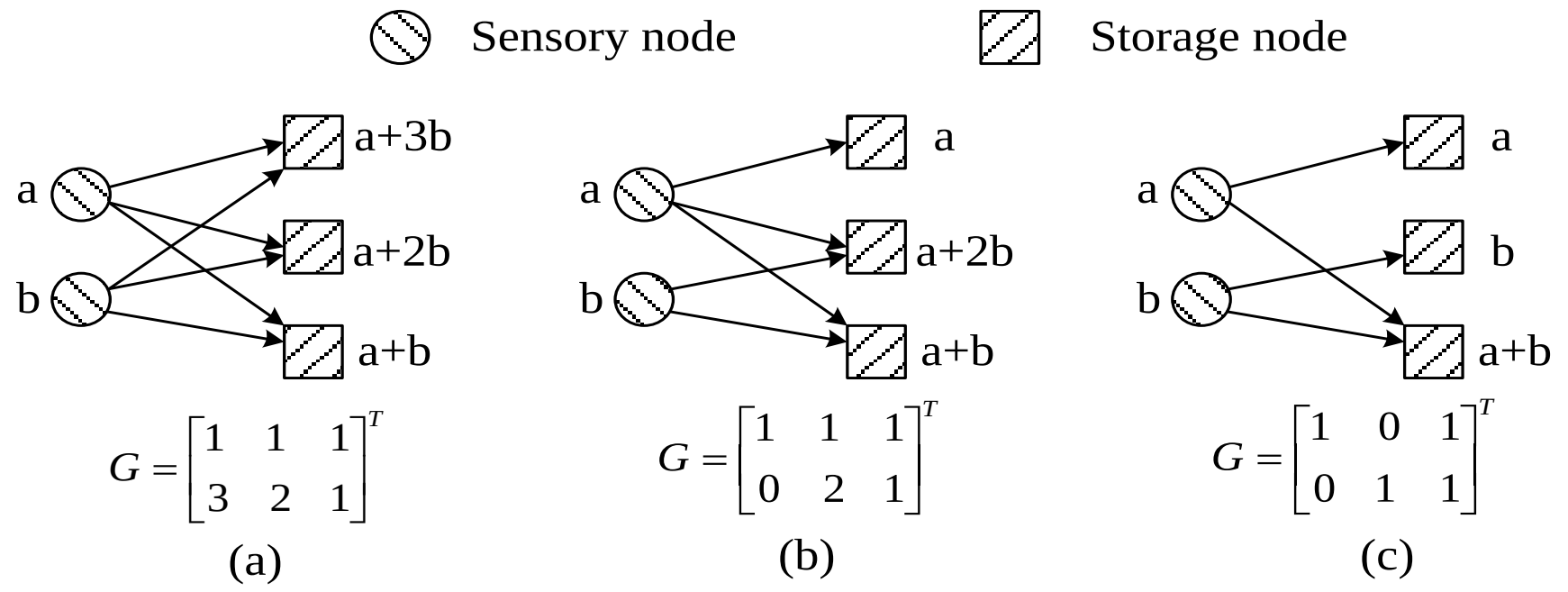

Figure 1 Relation between Energy Efficiency and Generator Matrix

It is straightforward to verify that all the above three diagrams have the (3, 2) MDS property that any two out of the three storage nodes suffice to recover. We can obviously observe that the sparser the generator matrix is, the fewer transmissions the network requires. Furthermore, the number of non-zero elements in the generator matrix equals the times of transmissions required by nodes in the examples. Therefore, the more zero elements in the generator matrix $G$, the higher energy efficiency the sensor network has.

In this section, we will provide a method to construct the sparse matrix. In addition, we will prove that the constructed matrix is the sparsest matrix with the MDS property, because a further sparser generator matrix will make it lose the MDS property. Therefore, we believe that our method can be used to obtain the theoretically sparsest generator matrix. This also means that after using the generator matrix obtained by this method, the network will consume the least energy to implement distributed storage.

We know that RS code is an MDS code, and most practical methods in distributed storage are related to this code. Through certain transformation of the generator matrix for RS code, our method can obtain the sparsest matrix for distributed storage.

Assume $M$ is an $n$-by-$k$ generator matrix for RS code, so $M$ has the MDS property. $N$ is a $k$-by-$k$ sub-matrix of matrix $M$, so the sparse generator matrix $G$ can be obtained through the following transformation:

$$G = MN^{-1} = \begin{bmatrix} a_1^1 & a_1^2 & \cdots & a_1^k \\ a_2^1 & a_2^2 & \cdots & a_2^k \\ \vdots & \vdots & \ddots & \vdots \\ a_k^1 & a_k^2 & \cdots & a_k^k \\ a_{k+1}^1 & a_{k+1}^2 & \cdots & a_{k+1}^k \\ \vdots & \vdots & \ddots & \vdots \\ a_n^1 & a_n^2 & \cdots & a_n^k \end{bmatrix} \begin{bmatrix} a_1^1 & a_1^2 & \cdots & a_1^k \\ a_2^1 & a_2^2 & \cdots & a_2^k \\ \vdots & \vdots & \ddots & \vdots \\ a_k^1 & a_k^2 & \cdots & a_k^k \end{bmatrix}^{-1} = \begin{bmatrix} 1 & 0 & \cdots & 0 \\ 0 & 1 & \cdots & 0 \\ \vdots & \vdots & \ddots & \vdots \\ 0 & 0 & \cdots & 1 \\ x_{(k+1)1} & x_{(k+1)2} & \cdots & x_{(k+1)k} \\ \vdots & \vdots & \ddots & \vdots \\ x_{n1} & x_{n2} & \cdots & x_{nk} \end{bmatrix} = \begin{bmatrix} I_{k\times k} \\ G'_{(n-k)\times k} \end{bmatrix} \quad (3.1)$$

We observe that the generator matrix $G$ is sparser than that of RS code because the generator matrix $G$ contains an identity matrix. Therefore, the sparsity has been improved. We must ensure that this matrix $G$ still has the MDS property, or this matrix makes no sense even though it has many zero elements.

***Theorem 1:*** If $M$ is an $n$-by-$k$ generator matrix with the MDS property and $Q$ is an $k$-by-$k$ full-rank matrix, the elements in both matrices $M$ and $Q$ are from $GF(q)$, then matrix $G = MQ$ still has the MDS property over the field $GF(q)$.

***Proof*:** For any $k$-by-$k$ sub-matrix $A$ formed by $k$ row vectors in matrix $G$, there must exist an $k$-by-$k$ sub-matrix $B$ of matrix $M$ such that $A = BQ$. Given that any $k$ row vectors in the matrix $M$ must form a full-rank matrix, so matrix $A$ must be a full-rank matrix. Therefore, matrix $G$ still has the MDS property, which implies that matrix $G$ can be employed as a generator matrix, and the codes constructed by this matrix $G$ are still MDS codes.

After proving this theorem, we can be sure that adoption of this sparse matrix can improve the energy consumption of WSNs. Now, the problem is whether we can make this matrix further sparser in order to achieve a higher energy-efficiency.

***Theorem 2:*** After the transformation shown in Eq. (3.1), the generator matrix $G$ is the sparsest matrix with the $(n,k)$ MDS property, namely, the number of zero elements reaches the maximum possible value.

***Proof*:** Firstly, it is obvious that the identity sub-matrix $I_{k\times k}$ can not be further sparser, or it will be singular. Then we address whether the matrix $G'_{(n-k)\times k}$ can become sparser. Assume that among these vectors, there is a vector $\alpha$ with a zero element. Without loss of generality, we assume that the $j^{th} (j \in [1,k])$ element of $\alpha$ is zero, which is shown as follows:

$$\alpha = [x_1 \quad x_2 \quad \cdots \quad x_{j-1} \quad 0 \quad x_{j+1} \quad \cdots \quad x_k] \quad (3.2)$$

Note that the first $k$ row vectors of $G$ form an identity matrix. If we replace the $j^{th}$ identity row vector of $I_{k\times k}$ with the vector $\alpha$, we will obtain the following square sub-matrix $S$.

$$S = \begin{bmatrix} 1 & & & & & & \\ & \ddots & & & & & \\ & & 1 & & & & \\ x_1 & \cdots & x_{j-1} & 0 & x_{j+1} & \cdots & x_k \\ & & & & 1 & & \\ & & & & & \ddots & \\ & & & & & & 1 \end{bmatrix} \quad (3.3)$$

According to ***Theorem 1***, matrix $G$ has the MDS property. However, we observe that matrix $S$ must be a singular matrix because the $j^{th}$ column vector is a zero vector, which violates ***Theorem 1***. Therefore, the assumption is untenable, and ***Theorem 2*** is proven. Then, we conclude that the number of zero elements has reached the maximum, which implies that the required transmissions are reduced to the minimum. Specifically, the total weight of these $k$ codewords is $k \times (n-k+1)$.

Then, we address the load balancing of these sensory nodes. Each column vector in matrix $G$ represents a code word. Moreover, the number of required transmission depends on the weights of the codewords. In accordance with ***Theorem 2***, we know that this matrix has reached its sparsest form, there couldn't be any zero elements in matrix $G'$, then we obtain the following ***Theorem 3***.

***Theorem 3****:* After the transformation shown in Eq. (3.1), each code word in the matrix $G$ must have the same weight.

***Proof:*** According to ***Theorem 2***, this theorem can be easily proved.

Therefore, from the perspective of load balancing, the energy consumption of sensory nodes will be highly balanced after using the generator matrix $G$.

## 4 Network coding-based iterative repair technology for WSNs

Due to unstable node performance in WSNs, the nodes tend to fail. When one or several nodes out of $n$ storage nodes in the network fail, the service level of the storage system will be reduced. Therefore, data of the failed nodes needs to be repaired at some new storage nodes to maintain the $(n,k)$ MDS property.

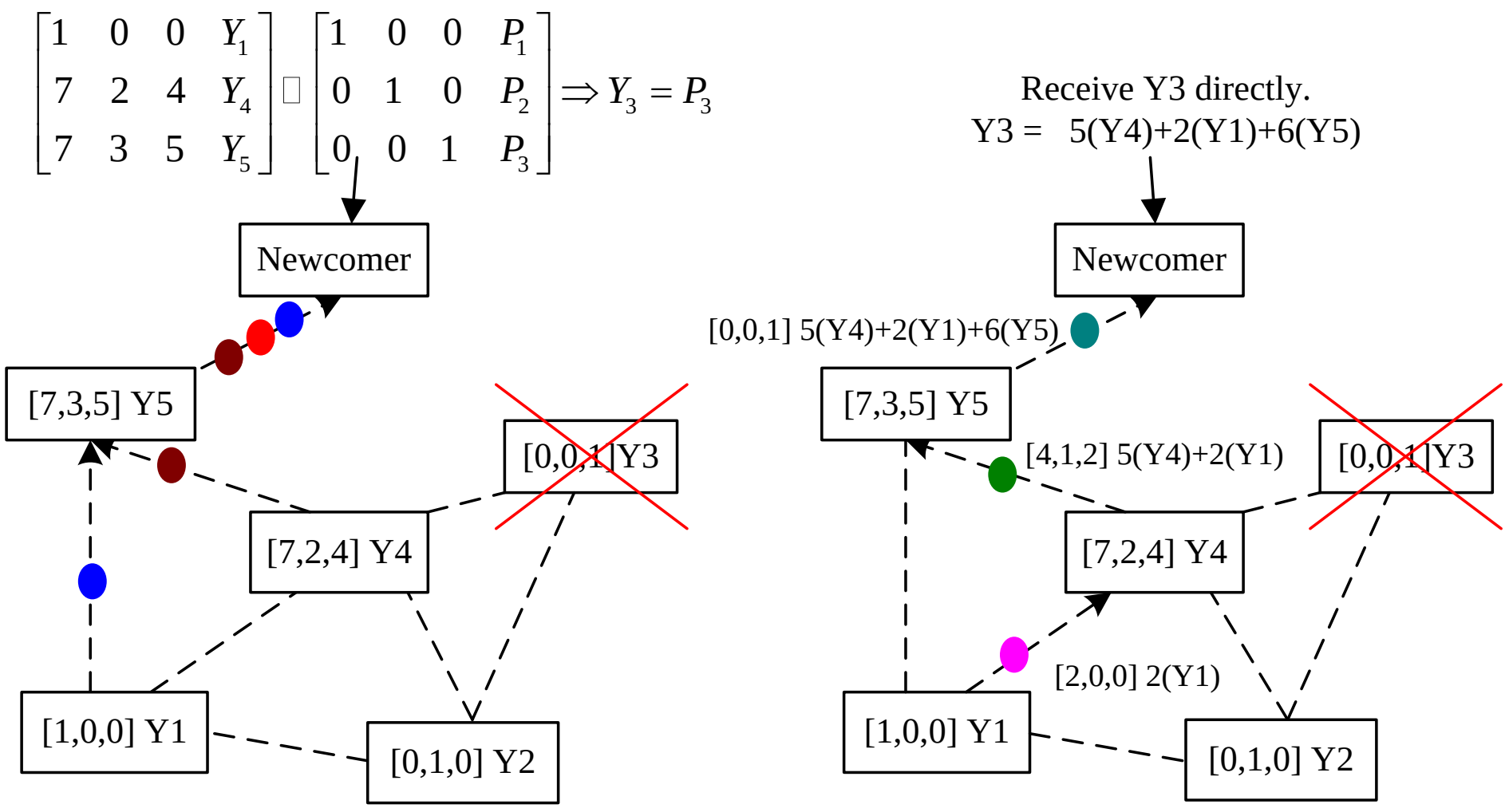

(a) Traditional repair technology (b) NC-based repair technology

Figure 2 Traditional Repair and NC-based Iterative Repair

In this example, (5, 3) MDS code is employed, so any 3 out of these 5 shares suffice to recover. First of all, we need to obtain a sparse generator matrix with the method in the Section 3. Then we will obtain following matrix $G$ for this example.

$$G = MN^{-1} = \begin{bmatrix} 1 & 4 & 5 & 6 & 7 \\ 1 & 2 & 3 & 4 & 5 \\ 1 & 1 & 1 & 1 & 1 \end{bmatrix}^T \begin{bmatrix} 1 & 1 & 1 \\ 1 & 2 & 4 \\ 1 & 3 & 5 \end{bmatrix}^{-1} = \begin{bmatrix} 1 & 0 & 0 & 4 & 5 \\ 0 & 1 & 0 & 2 & 3 \\ 0 & 0 & 1 & 7 & 7 \end{bmatrix}^T$$

In this example, the employed field is $GF(8)$, and the primitive polynomial is $D^3 + D$ which is employed to generate the elements in the field. According to the ***Theorem 2***, it is impossible to obtain a sparser generator matrix without losing the MDS property. Therefore, we employ the matrix $G$ to implement distributed storage, as shown in Fig. 2. When a storage node fails, we need to regenerate the data at a newcomer. Fig. 2(a) shows the traditional method: one new storage node must download at least $k$ sub files, then the original file will be recovered, and at last, new data will be regenerated. Then we introduce our network coding-base repair method.

### 4.1 Construction of optimal repair tree

In order to regenerate the data in the failed storage node, we should determine the surviving nodes participating in the repair process. In traditional repair method, the $k$ storage nodes closest to the newcomer are preferred to help regenerate the new data. The reason is that the closer the distance, the fewer hops are required, and therefore the repair process will have less energy overhead. Then, we will discuss after using the repair method based on network coding, which storage nodes should be chosen to participate in the repair process. We prefer to find the repair tree with the longest hops, and we will explain our preference with the following diagram.

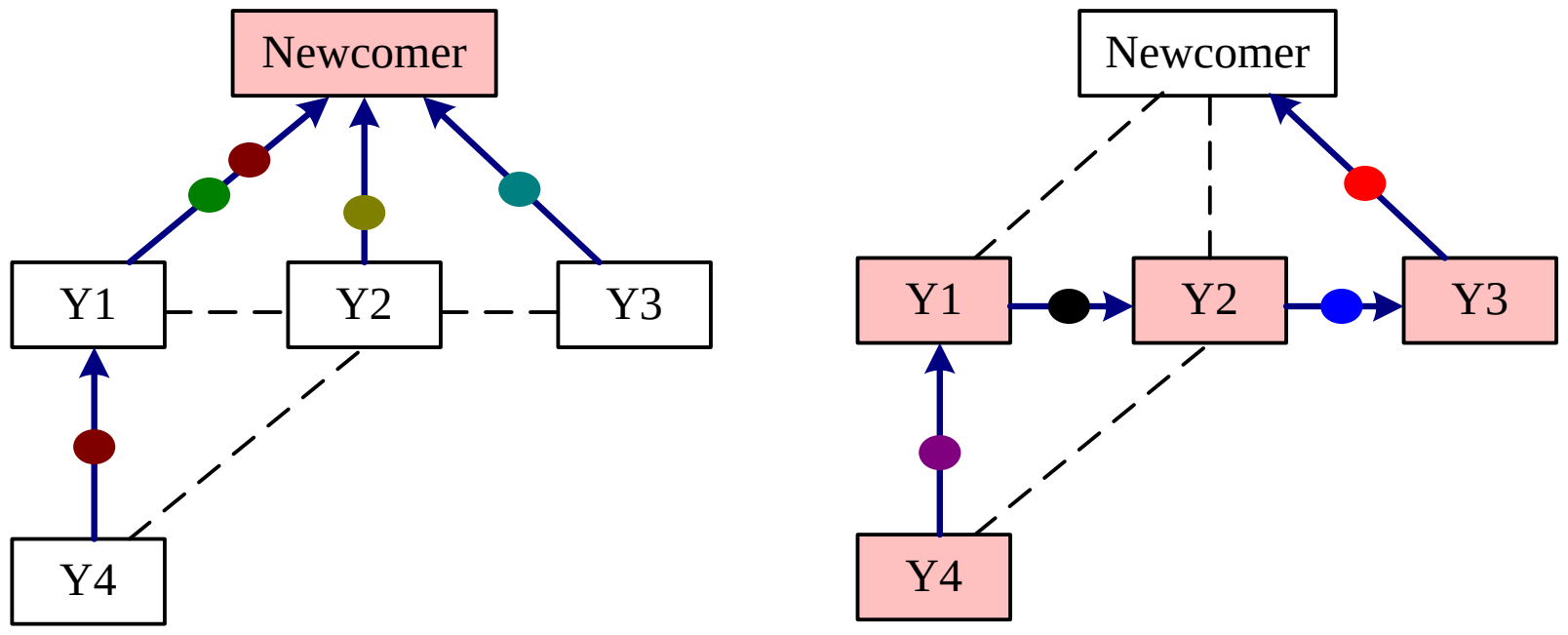

(a) Traditional Method (b) NC-based method

Figure 3 Repair Tree in Traditional Method and our NC-Based Method

The above diagram can well show our preference. Fig. 3(a) shows the traditional repair method, while Fig. 3(b) shows our repair method. In the two figures, red nodes mean that coding has been performed at these nodes. We prefer to construct a long repair tree since the network computation load of encoding will be highly balanced. In a general network environment, we employ the following algorithm to construct the repair tree.

**Algorithm 1:** Construction of optimal repair tree

```
// Stage 1: Repair notification
The new node broadcasts a RNC (Repair Tree Construction) packet to its neighbors.
In the RNC packet, an empty route sequence is included.
For each neighbor receiving the RNC packet
    If its ID is included in the route sequence.
        Gives up the RNC packet.
    Else
        Inserts its ID into the sequence, and forwards the packet.
        Records the longest route to the newcomer.
    End
End
```

**// Stage 2: Feedback**
**For** each neighbor
    Sends its longest route and the coding vector $\alpha$ to the new node
**End**
**// Stage 3: Construct the repair tree**
**If** there is a route with k storage nodes.
    The k nodes in the route will be selected to be repair tree.
**Else**
    Select the r nodes in the longest route, and the other k-r nodes which
    connect to the nodes in the longest route. These k nodes will form the tree.
**End**

In stage 1, the newcomer needs to have the knowledge of network topology so that it could calculate an effective repair scheme and initialize the repair process. Therefore, it needs to broadcast the RNC packets to gather necessary information. After each neighbor reports its longest route to the newcomer and coding vector, the newcomer will start to construct the repair tree. As mentioned above, this method prefers finding a long repair tree without any branch, so it will traverse all the routes it receives, and find a route consisting of $k$ storage nodes. Since each feedback packet includes a route sequence, the newcomer could determine the $k$ storage nodes according to this information. If the number of storage nodes in the longest route is still smaller than $k$, the longest route with $r < k$ nodes will be selected, and $k-r$ other nodes connected to the longest route will be selected. These $k$ storage nodes will form the repair tree, which is regarded as a suboptimum repair scheme.

We have to admit that our method has one shortage compared with the traditional method. Because we prefer to choose a collection of storage nodes that can form a long main tree, a long main path may cause long transmission delay. We believe that in distributed storage in WSNs, energy consumption balance of the system is an important factor since it determines the running time of the network, while transmission delay is not particularly important in such a system. After weighing the advantages and disadvantages, we prefer the former.

### 4.2 Regenerate the data

After the construction of repair tree, the $k$ storage nodes required to repair will be determined. Moreover, the newcomer will have the knowledge of the coding vector set of these $k$ storage nodes $[\alpha_1, \alpha_2, ..., \alpha_k]$ in which each element represents a $k$-dimensional vector. Then, we need to allocate a coding vector for the new storage node. As mentioned, we employ a sparse generator matrix to generate network codes. If we choose one vector that has not been used from this matrix, then this method will be regarded as functional repair [19]; if the coding vector used by failed node is allocated again, this method will be regarded as exact repair. No matter the repair mode is functional repair or exact repair, the $k$-dimensional target coding vector $\beta = [b_1, b_2, ..., b_k]$ will be obtained. In the example shown in Fig. 2, after using network coding, we need to solve a group of coefficients to enable the intermediate nodes to re-encode. In general cases, we need to obtain the $k$ values of $(x_1, x_2, ..., x_k)$, and these $k$ values must satisfy that the equation $\beta = x_1\alpha_1 + x_2\alpha_2 + ... + x_k\alpha_k$ holds. These $k$ values can be easily obtained by solving a system of linear equations. Obviously, the group of data $(x_1, x_2, ..., x_k)$ must exist since the matrix $[\alpha_1, \alpha_2, ..., \alpha_k]^T$ is a $k$ -by- $k$ sub-matrix of the generator matrix $G$, which must be a full-rank matrix. After obtaining these $k$ coefficients, some control packets will be sent to these $k$ storage nodes to notify them how to re-encode during the repair process. After receiving the control packets, the storage nodes at the tail of the tree will encode their stored data and send it to the next hop, each intermediate node participating in repair process needs to re-encode, and finally, the new storage node will receive the regenerated data.

The repair method can be easily extended to multiple failures of storage nodes. When there are multiple failures, the repair process should be performed for the same number of times. When a newcomer has repaired the data in a failed node, it can serve as a surviving node to participate in further repair. Although multiple failures can be repaired by repeatedly using the repair method, there is a precondition that the number of failures should not exceed $(N-K)$ when $(N, K)$ MDS code is employed so that the surviving storage nodes suffice to decode and repair.

### 4.3 Advantages of network coding-based iterative data repair

In the previous section, we have provided the repair scheme, and then we address the advantages of the proposed iterative repair technology. Compared with traditional recovery technology, the recovery technology which uses network coding in WSNs mainly has the following advantages:

**(1) Energy efficiency advantage**

Fig.2(a) shows the traditional recovery method, and the newcomer needs to download at least 3 code blocks to repair the missed block. This problem can be improved when network coding is employed. After using network coding, the newcomer only needs to download 1 code block, as shown in Fig.2(b). Because nodes in the WSNs self-organize the network in a multi-hop way, the traditional method requires 5 data transmissions to complete the repair process. In Fig.2(b), the re-encoding operation from network coding theory is adopted, and then we observe that only 3 transmissions are required, which means energy consumption of the network will be reduced. Note that the re-encoding operation is no longer employed to increase throughput, but to reduce the repair bandwidth. Therefore, the re-encoding operation is different in function from that in traditional multicast networks, but the core idea is the same, namely, re-encoding operation is performed at intermediate nodes to achieve performance improvement.

**(2) Load balancing advantage**

The advantage of load balancing is mainly reflected on two aspects: the transmission load and the computation load. In accordance with Fig. 2, we observe that when network coding is not adopted, the network load is severely unbalanced, and the storage node closer to the newcomer needs to consume more energy, because it does not only have to transmit its own stored data to the newcomer, but also the data sent from upstream nodes to the newcomer, so load of this node is much higher than others. After using network coding, each storage node on the repair path only has to transmit once, so each storage node has equal energy consumption.

When data in the network has a big size, the computation overhead should not be neglected. Traditional method can be considered as a centralized method, while our method is a decentralized method. In the traditional method, all of the computation overhead is concentrated on the newcomer. After adoption of network coding, all the intermediate nodes on the repair path need to participate into the computation of new data blocks. In this way, the computation overhead will be equally distributed. Moreover, after using network coding for data repair, the data received by the newcomer is the very data that needs to be stored.

**(3) Buffer overhead and security advantages**

We notice that in the traditional scheme, the newcomer must have adequate buffer space to store the $k$ received sub-files, while the repair scheme based on network coding has much lower requirement of buffer space. The buffer space of 1 sub-file is enough.

The repair strategy shown in Fig.2(b) is a decentralized one, and during the repair process, this strategy makes each participating node unable to have all the original packets. However, in traditional method, the newcomer and its adjacent nodes will receive all the $k$ sub-files, which suggests that when these nodes are occupied by an eavesdropper, it will be able to steal all the original data. The decentralized iterative repair strategy based on network coding can ensure that during the repair process, the original data won't converge to a certain node, which will definitely increase the system security.

**(4) Additional advantage**

In a common distributed storage system, each storage node generally only stores one data, so it requires $n$ storage nodes to store $n$ encoded data. Dimakis et al. [19] proposed an interesting example to demonstrate the advantages of the distributed storage system based on network coding. A significant feature of the example is: for $n$ encoded data, it is allowed to store multiple data at the same storage node. During the repair process, at least $k$ data are also required to participate into the repair process, but because some nodes have multiple encoded data, these data blocks can be operated within the same storage node without additional transmission, so then the energy consumption of transmission will be reduced. Although our NC-based repair method is different from the previous NC-based repair method, this idea could be compatible with our method, which is explained by the following example.

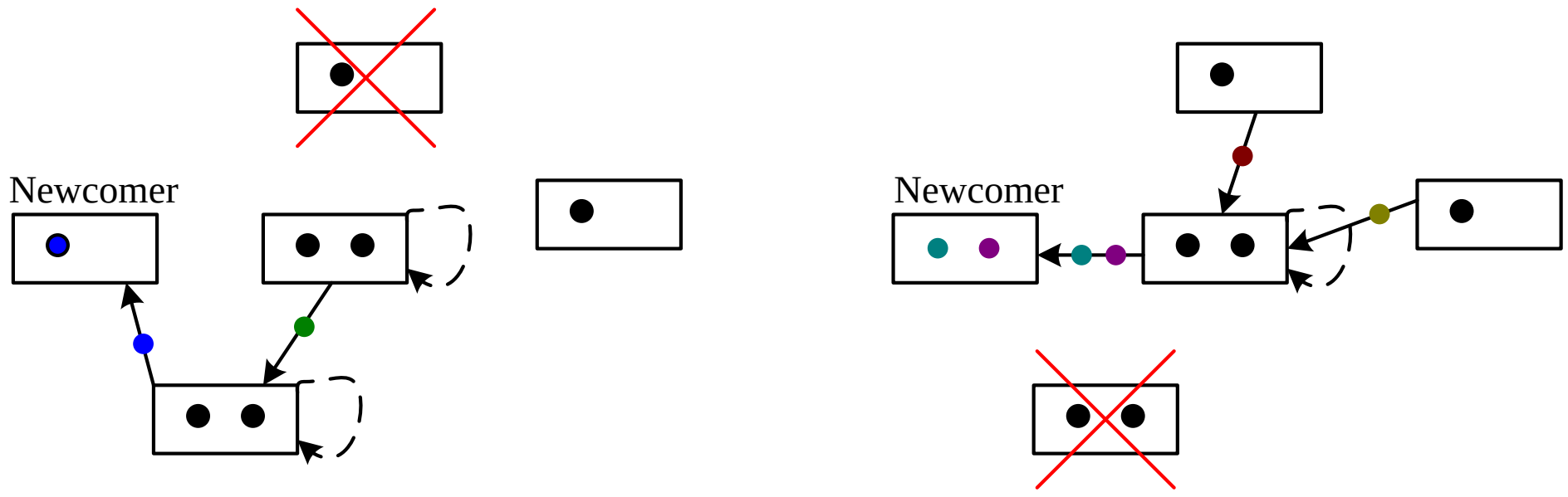

(a) Storage node with 1 data fails (b) Storage node with 2 data fails

Figure 4 Benefit of data repair based on network coding

In Fig. 4, the original file is encoded into six sub-files, and any four out of the six sub-files suffice to recover the original file. However, there are only four storage servers in the initial state, so two servers are allowed to store two sub-files. When a storage server fails, it is possible to regenerate a new sub-file with the bandwidth of two sub-files, as shown in Fig.4(a). Therefore, the performance of network coding is better reflected under this circumstance, since the storage nodes can re-encode the multiple sub-files without much more transmissions, so the energy efficiency is further improved. Even if the storage node with two sub-files fails, only 4 sub-files need to be transmitted, as shown in Fig. 4(b). If the traditional repair method is employed, 6 sub-files need to be transmitted. It should be noted that after storing multiple data in a storage node, the reliability will be reduced since one failure will result in loss of multiple sub-files, which is inevitable sometimes. For example, when there isn't enough storage servers (less than $n$) to store the $n$ encoded sub-files, part of the storage nodes should be allowed to store more than 1 sub-file. Another example is that, when a storage node fails, we need to reconstruct the data on a new storage node. However, there may be no vacant storage node. In this case, repairing the missing data on an occupied storage node is the only way, which will also lead to multiple sub-files storing in one storage node.

Overall, when it is allowed to store multiple sub-block data in one storage node, the repair strategy based on network coding can further reduce the required bandwidth and energy during the repair process.

## 5 Required Minimum Size of the Finite Field

Theoretically, the computation of network coding over a small finite field is more energy-efficient than that over a large finite field. Therefore, we will address the lower bound of the required finite field in this section, and we will provide an explicit lower bound for our scheme.

***Theorem 4:*** For an $n$-by-$k$ generator matrix for MDS code, the field $GF(q)$ with a size of $q=n-1,(k<q)$ is sufficient, unless $n=q+2=2^m+2$ and $k=3$ or $k=q-1$, in which case $q=n-2$ is sufficient.

***Proof:*** As long as we prove that, over $GF(q)$, at least $q+1$ vectors can be obtained for the generator matrix and $q+2$ in the special cases, the ***Theorem 4*** will be proven.

We start with the Vandermonde matrix, it is obvious that $q-1$ different nonzero elements can be selected from the given field $GF(q)$ to construct the row vectors $[1,t,t^2,...,t^{k-1}](t\in F_q,t\neq 0)$. These $q-1$ vectors can be employed to form the generator matrix since the determinate of a Vandermonde matrix is nonzero. For convenience, we denote the vector space formed by the $q-1$ vectors as $V$, and we address whether some other vectors can be added to $V$ such that the MDS property is still maintained.

It is straightforward to prove that the identity vectors $e_1=(1,0,...,0)$ and $e_k=(0,0,...,1)$ can be added to vector space $V$.

$$S=\begin{bmatrix} 1 & 0 & \cdots & 0 & 0 \\ a_1^0 & a_1^1 & \cdots & a_1^{k-2} & a_1^{k-1} \\ \vdots & \vdots & \ddots & \vdots & \vdots \\ a_{k-2}^0 & a_{k-2}^1 & \cdots & a_{k-2}^{k-2} & a_{k-2}^{k-1} \\ a_{k-1}^0 & a_{k-1}^1 & \cdots & a_{k-1}^{k-2} & a_{k-1}^{k-1} \end{bmatrix} \rightarrow \begin{bmatrix} 1 & 0 & \cdots & 0 & 0 \\ 0 & a_1^1 & \cdots & a_1^{k-2} & a_1^{k-1} \\ \vdots & \vdots & \ddots & \vdots & \vdots \\ 0 & a_{k-2}^1 & \cdots & a_{k-2}^{k-2} & a_{k-2}^{k-1} \\ 0 & a_{k-1}^1 & \cdots & a_{k-1}^{k-2} & a_{k-1}^{k-1} \end{bmatrix} \tag{5.1}$$

In Eq.(5.1), $a_1,a_2,...,a_{k-1}$ are different nonzero elements from $GF(q)$. Then, the $(k-1)$-by-$(k-1)$ sub-matrix in the frame must be a full-rank matrix since it is equivalent to a Vandermonde matrix. Therefore, $e_1=(1,0,...,0)$ can be added to vector space $V$. Similarly, $e_k=(0,0,...,1)$ can be added to $V$ as well. Moreover, $e_1$ and $e_k$ can be simultaneously added to $V$ since the $(k-2)$-by-$(k-2)$ sub-matrix shown in Eq.(5.2) is also equivalent to a Vandermonde matrix.

$$S=\begin{bmatrix} 1 & 0 & \cdots & 0 & 0 \\ a_1^0 & a_1^1 & \cdots & a_1^{k-2} & a_1^{k-1} \\ \vdots & \vdots & \ddots & \vdots & \vdots \\ a_{k-2}^0 & a_{k-2}^1 & \cdots & a_{k-2}^{k-2} & a_{k-2}^{k-1} \\ 0 & 0 & \cdots & 0 & 1 \end{bmatrix} \rightarrow \begin{bmatrix} 1 & 0 & \cdots & 0 & 0 \\ 0 & a_1^1 & \cdots & a_1^{k-2} & 0 \\ \vdots & \vdots & \ddots & \vdots & \vdots \\ 0 & a_{k-2}^1 & \cdots & a_{k-2}^{k-2} & 0 \\ 0 & 0 & \cdots & 0 & 1 \end{bmatrix} \tag{5.2}$$

Then we prove that identity vector $e_{k-1}=(0,\ldots,1,0)$ can be added to $V$ when $q=2^m$, and $k=3$ or $k=q-1$.

$$\begin{bmatrix} a_1^0 & \cdots & a_1^{k-3} & a_1^{k-2} & a_1^{k-1} \\ a_2^0 & \cdots & a_2^{k-3} & a_2^{k-2} & a_2^{k-1} \\ \vdots & \ddots & \vdots & \vdots & \vdots \\ a_{k-1}^0 & \cdots & a_{k-1}^{k-3} & a_{k-1}^{k-2} & a_{k-1}^{k-1} \\ 0 & \cdots & 0 & 1 & 0 \end{bmatrix} \rightarrow \begin{bmatrix} a_1^0 & \cdots & a_1^{k-3} & 0 & a_1^{k-1} \\ a_2^0 & \cdots & a_2^{k-3} & 0 & a_2^{k-1} \\ \vdots & \ddots & \vdots & \vdots & \vdots \\ a_{k-1}^0 & \cdots & a_{k-1}^{k-3} & 0 & a_{k-1}^{k-1} \\ 0 & \cdots & 0 & 1 & 0 \end{bmatrix} \rightarrow \begin{bmatrix} a_1^0 & \cdots & a_1^{k-3} & a_1^{k-1} & 0 \\ a_2^0 & \cdots & a_2^{k-3} & a_2^{k-1} & 0 \\ \vdots & \ddots & \vdots & \vdots & \vdots \\ a_{k-1}^0 & \cdots & a_{k-1}^{k-3} & a_{k-1}^{k-1} & 0 \\ 0 & \cdots & 0 & 0 & 1 \end{bmatrix} \quad (5.3)$$

Note that the $(k-1)$-by-$(k-1)$ sub-matrix in the frame is a jump Vandermonde matrix. Then we employ the following method to check whether it is full rank.

$$\det(\Gamma)=\det(\begin{bmatrix} a_1^0 & \cdots & a_1^{k-3} & a_1^{k-2} & a_1^{k-1} \\ a_2^0 & \cdots & a_2^{k-3} & a_2^{k-2} & a_2^{k-1} \\ \vdots & \ddots & \vdots & \vdots & \vdots \\ a_{k-1}^0 & \cdots & a_{k-1}^{k-3} & a_{k-1}^{k-2} & a_{k-1}^{k-1} \\ x^0 & \cdots & x^{k-3} & x^{k-2} & x^{k-1} \end{bmatrix}) \quad (5.4)$$
$$=(\prod_{i\in[1,k-1]}(x-a_i))(\prod_{m<n,m,n\in[1,k-1]}(a_n-a_m))=\mathrm{X}(\prod_{i\in[1,k-1]}(x-a_i))$$

On the other hand, the determinate can also be expressed by the following equation.

$$\det(\Gamma)=(-1)^{k+1}(A_{k1}\times x^0-\ldots+A_{k2}\times x^{k-3}-\det(S)\times x^{k-2}+A_{kk}\times x^{k-1}) \quad (5.5)$$

Therefore, the coefficient of $x^{k-2}$ in Eq. (5.5) must equal to that in Eq. (5.4). Then we need to check the coefficient of $x^{k-2}$ in Eq. (5.4), and if the coefficient identically equals to nonzero no matter what the elements $a_1,a_2,\ldots,a_{k-1}$ are, $e_{k-1}$ can be added to $V$. We know that $a_1,a_2,\ldots,a_{k-1}$ are different elements from the same field $GF(q)$, so $\mathrm{X}\neq 0$ when $q=2^m$. Then we can directly remove it since it won't affect checking whether the coefficient is zero or not.

$$\prod_{i\in[1,k-1]}(x-a_i)=(x-a_1)(x-a_2)\ldots(x-a_{k-1}) \quad (5.6)$$

In accordance with Eq. (5.6), the coefficient of $x^{k-2}$ is $C=a_1+a_2+\ldots+a_{k-1}$. Therefore, checking whether the jump Vandermonde matrix is non-singular is equivalent to checking whether $C$ is nonzero. When $C$ identically equals to nonzero, the jump Vandermonde matrix must be nonsingular.

When $k=3$, $C=a_1+a_2$. When $q=2^m$, the addition operation is implemented with "XOR", $a_1$ and $a_2$ are different elements from the $GF(q)$, so $C$ must be nonzero.

When $k=q-1$, $C=a_1+a_2+\ldots+a_{q-2}$. We know that when $q=2^m$ the sum of all the $q-1$ nonzero elements in $GF(q)$ must be 0, and $C$ is the sum of $q-2$ different elements, so $C$ must be nonzero.

Moreover, we can conclude that when $k=q-2$, $C$ must be nonzero as well. A brief proof is as follows. Because the sum of any two nonzero elements in $GF(q)$ must be nonzero and the sum of all the $q-1$ nonzero elements must be zero, the sum of any $q-3$ nonzero elements must be nonzero. Therefore, when $k=q-2$, $C$ must be nonzero.

In conclusion, when $q=2^m$ and $k=3$ or $k=q-2$ or $k=q-1$, $e_{k-1}$ can be added to $V$.

We have proven that $e_1$ and $e_k$ can be simultaneously added to $V$. Then, we need to address whether $e_1$, $e_{k-1}$ and $e_k$ can be simultaneously added to $V$.

$$\begin{bmatrix} a_1^0 & \cdots & a_1^{k-3} & a_1^{k-2} & a_1^{k-1} \\ \vdots & \ddots & \vdots & \vdots & \vdots \\ a_{k-2}^0 & \cdots & a_{k-2}^{k-3} & a_{k-2}^{k-2} & a_{k-2}^{k-1} \\ 0 & \cdots & 0 & 1 & 0 \\ 0 & \cdots & 0 & 0 & 1 \end{bmatrix} \to \begin{bmatrix} a_1^0 & \cdots & a_1^{k-3} & 0 & 0 \\ \vdots & \ddots & \vdots & \vdots & \vdots \\ a_{k-2}^0 & \cdots & a_{k-2}^{k-3} & 0 & 0 \\ 0 & \cdots & 0 & 1 & 0 \\ 0 & \cdots & 0 & 0 & 1 \end{bmatrix} \tag{5.7}$$

Because the sub-matrix in the frame is a Vandermonde matrix, $e_{k-1}$ and $e_k$ can be simultaneously added to $V$ when $q = 2^m$ and $k = 3$ or $k = q - 2$ or $k = q - 1$.

Then we consider whether $e_1$ and $e_{k-1}$ can be simultaneously added to $V$.

$$\begin{bmatrix} a_1^0 & a_1^1 & \cdots & a_1^{k-3} & a_1^{k-2} & a_1^{k-1} \\ a_2^0 & a_2^1 & \cdots & a_2^{k-3} & a_2^{k-2} & a_2^{k-1} \\ \vdots & \vdots & \ddots & \vdots & \vdots & \vdots \\ a_{k-2}^0 & a_{k-2}^1 & \cdots & a_{k-2}^{k-3} & a_{k-2}^{k-2} & a_{k-2}^{k-1} \\ 0 & 0 & \cdots & 0 & 1 & 0 \\ 1 & 0 & \cdots & 0 & 0 & 0 \end{bmatrix} \to \begin{bmatrix} 0 & a_1^1 & \cdots & a_1^{k-3} & 0 & a_1^{k-1} \\ 0 & a_2^1 & \cdots & a_2^{k-3} & 0 & a_2^{k-1} \\ \vdots & \vdots & \ddots & \vdots & \vdots & \vdots \\ 0 & a_{k-2}^1 & \cdots & a_{k-2}^{k-3} & 0 & a_{k-2}^{k-1} \\ 0 & 0 & 0 & 0 & 1 & 0 \\ 1 & 0 & 0 & 0 & 0 & 0 \end{bmatrix} \to \begin{bmatrix} 0 & a_1^1 & \cdots & a_1^{k-3} & a_1^{k-1} & 0 \\ 0 & a_2^1 & \cdots & a_2^{k-3} & a_2^{k-1} & 0 \\ \vdots & \vdots & \ddots & \vdots & \vdots & \vdots \\ 0 & a_{k-2}^1 & \cdots & a_{k-2}^{k-3} & a_{k-2}^{k-1} & 0 \\ 0 & 0 & 0 & 0 & 0 & 1 \\ 1 & 0 & 0 & 0 & 0 & 0 \end{bmatrix} \tag{5.8}$$

In the equation, $a_1, a_2, ..., a_{k-2}$ could be any $k-2$ out of the nonzero elements in $GF(q)$. When $k = 3$, the transformation is as follows.

$$\begin{bmatrix} a_1^0 & a_1^1 & a_1^2 \\ 0 & 1 & 0 \\ 1 & 0 & 0 \end{bmatrix} \to \begin{bmatrix} 0 & 0 & a_1^2 \\ 0 & 1 & 0 \\ 1 & 0 & 0 \end{bmatrix} \tag{5.9}$$

Obviously, the above matrix is a full-rank matrix when $k = 3$ and $q = 2^m$.

When $k = q - 1$, if the matrix in the above frame is full rank, $e_1$ and $e_{k-1}$ can be simultaneously added to $V$. Interestingly, the matrix in the frame is a jump Vandermonde matrix with the dimension of $k = q - 2$, and we have proven that it must be full rank in the above. Therefore, $e_1$ and $e_{k-1}$ can be simultaneously added to $V$ when $q = 2^m$ and $k = q - 1$. Similarly, if this conclusion holds for $k = q - 2$, a precondition that $e_{k-1}$ must be able to be added to $V$ when $k = q - 3$ should be satisfied. However, we don't have a conclusion to satisfy this precondition.

Therefore, when $k = 3$ and $k = q - 1$, $e_1$ and $e_{k-1}$ can be simultaneously added to $V$.

At last, we address whether $e_1$, $e_{k-1}$, $e_k$ can be simultaneously added to the generator matrix when $q = 2^m$ and $k = 3$ or $k = q - 1$.

When $k = 3$, this proposition is obviously true since the matrix is a 3-by-3 identity matrix which must be full rank. When $k = q - 1$, the proof is as follows.

$$\begin{bmatrix} a_1^0 & a_1^1 & \cdots & a_1^{k-3} & a_1^{k-2} & a_1^{k-1} \\ \vdots & \vdots & \ddots & \vdots & \vdots & \vdots \\ a_{k-3}^0 & a_{k-3}^1 & \cdots & a_{k-3}^{k-3} & a_{k-3}^{k-2} & a_{k-3}^{k-1} \\ 1 & 0 & \cdots & 0 & 0 & 0 \\ 0 & 0 & \cdots & 0 & 1 & 0 \\ 0 & 0 & \cdots & 0 & 0 & 1 \end{bmatrix} \to \begin{bmatrix} 0 & a_1^1 & \cdots & a_1^{k-3} & 0 & 0 \\ \vdots & \vdots & \ddots & \vdots & \vdots & \vdots \\ 0 & a_{k-3}^1 & \cdots & a_{k-3}^{k-3} & 0 & 0 \\ 1 & 0 & \cdots & 0 & 0 & 0 \\ 0 & 0 & \cdots & 0 & 1 & 0 \\ 0 & 0 & \cdots & 0 & 0 & 1 \end{bmatrix} \tag{5.10}$$

Obviously, the sub-matrix in the frame is equivalent to a Vandermonde matrix, so $e_1$, $e_{k-1}$ and $e_k$ can be simultaneously added to $V$ when $q = 2^m$ and $k = 3$ or $k = q - 1$. Therefore, there are in total $q+2$ row vectors in the special cases.

The MDS main conjecture [21] implies that the maximum length $n$ of MDS code may be $q+1$, and $q+2$ in the special cases. In this section, we prove that the maximum length $n$ is at least $q+1$, and $q+2$ in the special cases, and this result enables us to conclude that $q=n-1$ is sufficient to implement $(n,k)$ MDS code, and when $q=2^m$ and $k=3$ or $k=q-1$, $q=n-2$ is sufficient.

We have shown that a very small finite field is sufficient to implement this scheme, so the computation overhead of network coding can be low.

## 6 Simulations and analysis

We used Omnet++ [22] to implement the scheme, and a network topology was simulated. The simulation topology consisted of 10 storage nodes and 20 sensory nodes, both randomly deployed in a 200m*180m square area. The transmission radius of the each node is 60 meters. Moreover, IEEE 802.11b was used as the MAC layer protocol, and the channel bandwidth was set to be 2Mbps.

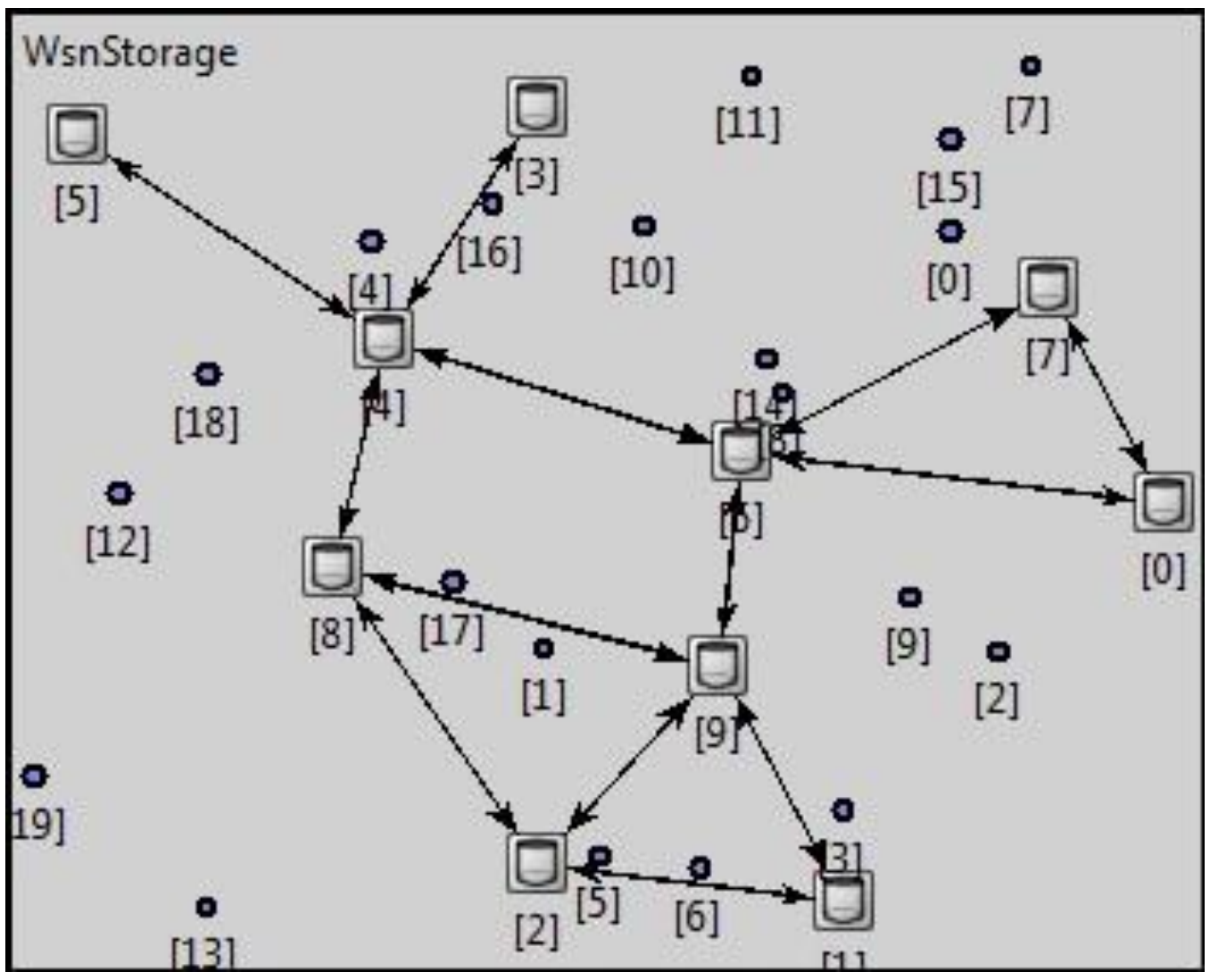

Figure 5 Simulated Wireless Sensor Networks

Based on the simulated network, we evaluated the energy efficiencies of the optimal storage technology in Section 3 and the repair technology in Section 4, respectively.

### 6.1 The performance of the proposed storage technology

**(1) Total energy consumption**

In WSNs, energy consumption mainly depends on the numbers of times to send and receive data. Therefore, when we evaluate the total energy consumption of the distributed storage system in WSNs, we actually evaluate the required data transmissions to implement the distributed redundant storage.

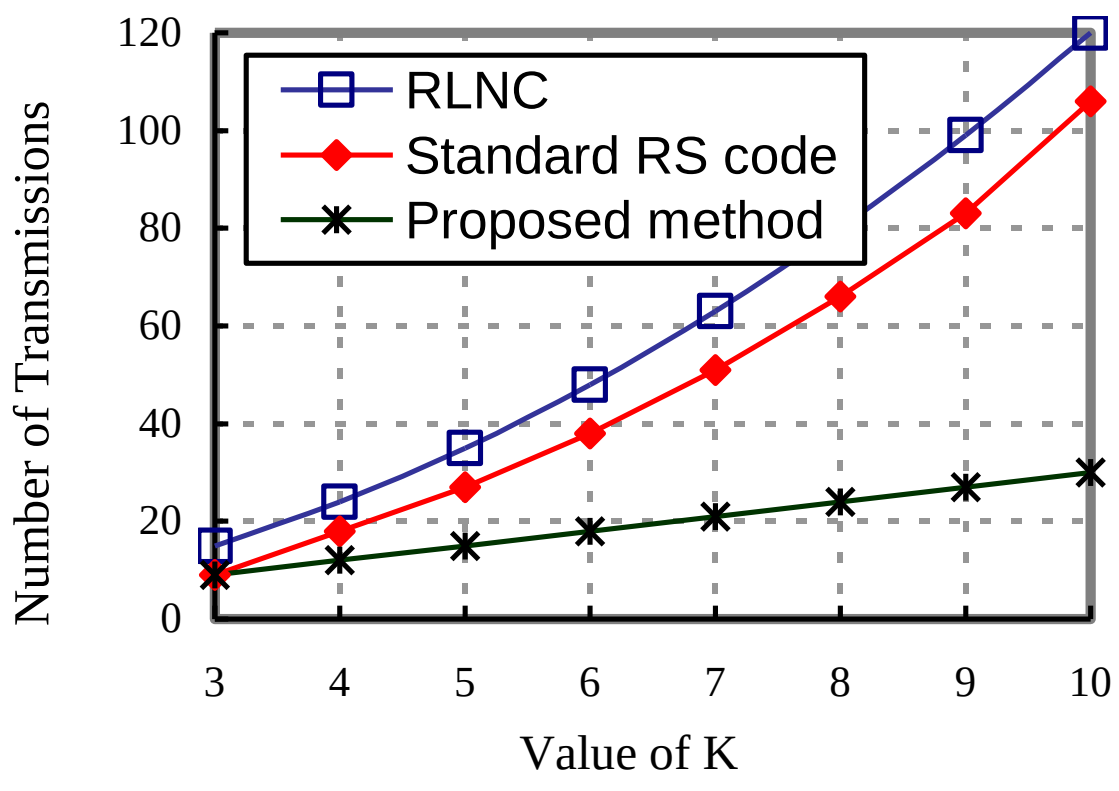

Figure 6 Energy Efficiency of Different Methods

In the simulations, data from $k$ sensor nodes will be encoded and sent to $n = k + 2$ storage nodes, and $(k+2, k)$ MDS property were maintained at these storage nodes. In Fig. 6, we observe that RLNC (Random Linear Network Coding) consumes the most energy, and the adoption of Reed-Solomon coding can improve the total energy consumption. In the simulation, when $K$ was greater than 3, we employed the following generator matrix $G_1$, and when $K$ equaled 3, we employed the following generator matrix $G_2$.

$$G_1 = \begin{bmatrix} 1 & a_1^0 & a_2^0 & \cdots & a_{n-3}^0 & 0 \\ \vdots & \vdots & \vdots & \ddots & \vdots & \vdots \\ 0 & a_1^{k-2} & a_2^{k-2} & \cdots & a_{n-3}^{k-2} & 0 \\ 0 & a_1^{k-1} & a_2^{k-1} & \cdots & a_{n-3}^{k-1} & 1 \end{bmatrix}^T, \quad G_2 = \begin{bmatrix} 1 & a_1^0 & \cdots & a_{n-4}^0 & 0 & 0 \\ 0 & a_1^1 & \cdots & a_{n-4}^1 & 1 & 0 \\ 0 & a_1^2 & \cdots & a_{n-4}^2 & 0 & 1 \end{bmatrix}^T$$

By using the method proposed in this paper, the network consumes the least energy. In particular, when $k = 3$, $e_1, e_2, e_3$ can be simultaneously added to the vector space $V$, as proven in Section 5. Therefore, using our method is equivalent to using RS code in this special case.

**(2) Load balancing of the network**

Energy load balancing is another important indicator to measure the energy efficiency. We have pointed out that the energy consumption required by each node is proportional to the weight of corresponding code word. In addition, we have proven that in the generator matrix obtained by using the proposed method, each column of the matrix represents a codeword, and the $k$ codewords must have equal weights. Therefore, by using this method, the balancing situation will reach theoretical optimum. Although RLNC has the worst performance in total energy consumption, the level of its load balancing is close to the method proposed in this paper. Because the elements of each code are obtained randomly, therefore, the weight of each code has equal mathematical expectation. It has the worst load balancing when RS codes are employed, because not all codes have the same weight. We have calculated the standard deviations of the simulation results, which are shown as follows.

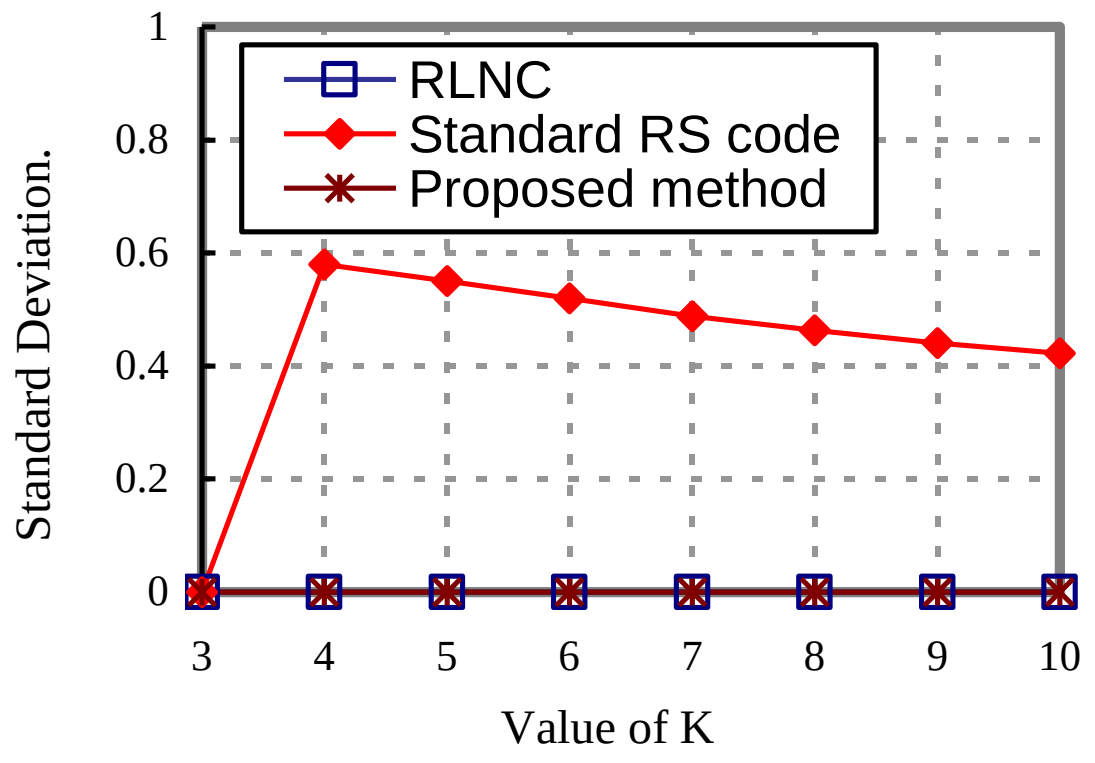

Figure 7 Load Balancing of Different Methods

As mentioned above, when $k=3$, the method which uses RS code is equivalent to our method, so, the load in this case can also be highly balanced.

**6.2 The performance of the proposed repair technology**

In order to evaluate the iterative repair method based on network coding, we need to implement the repair tree construction algorithm. Base on the simulated network shown in Fig. 5, we implement the proposed algorithm and the traditional algorithm for repair tree construction. For clarity, we remove the sensor nodes from the network during this stage.

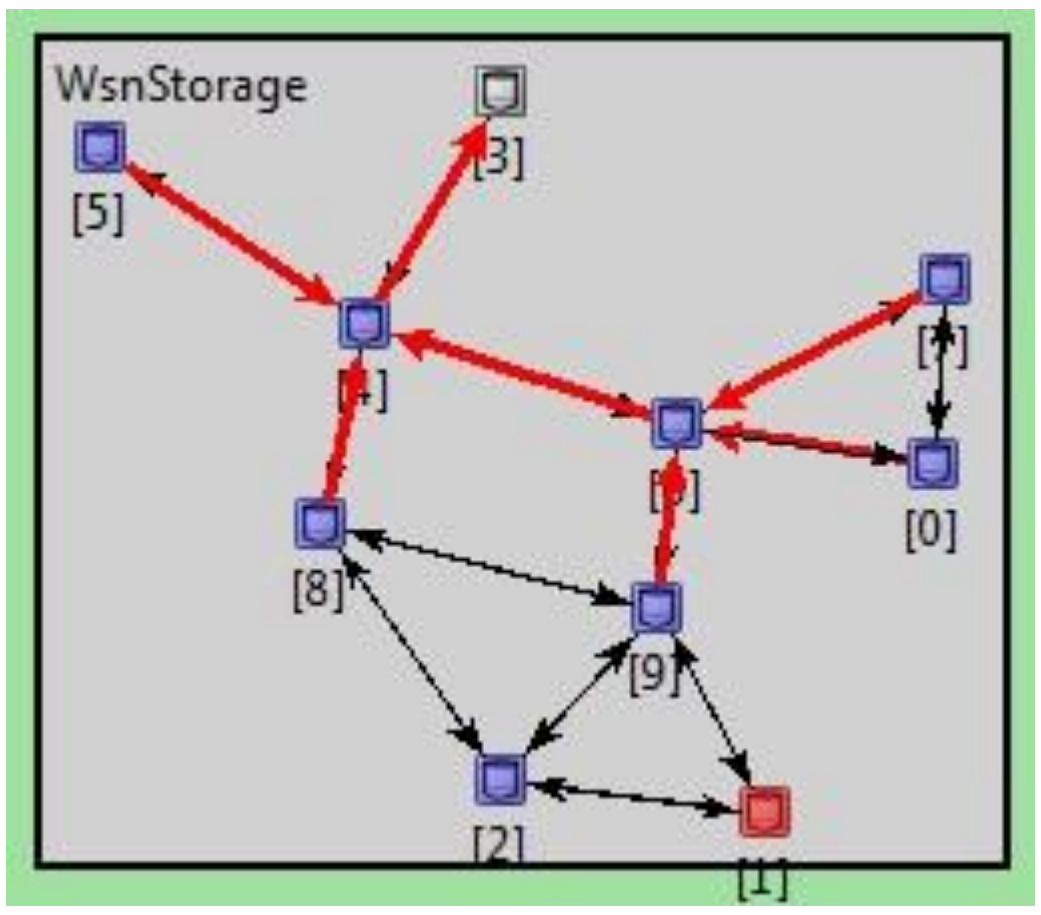

(a) Traditional repair method

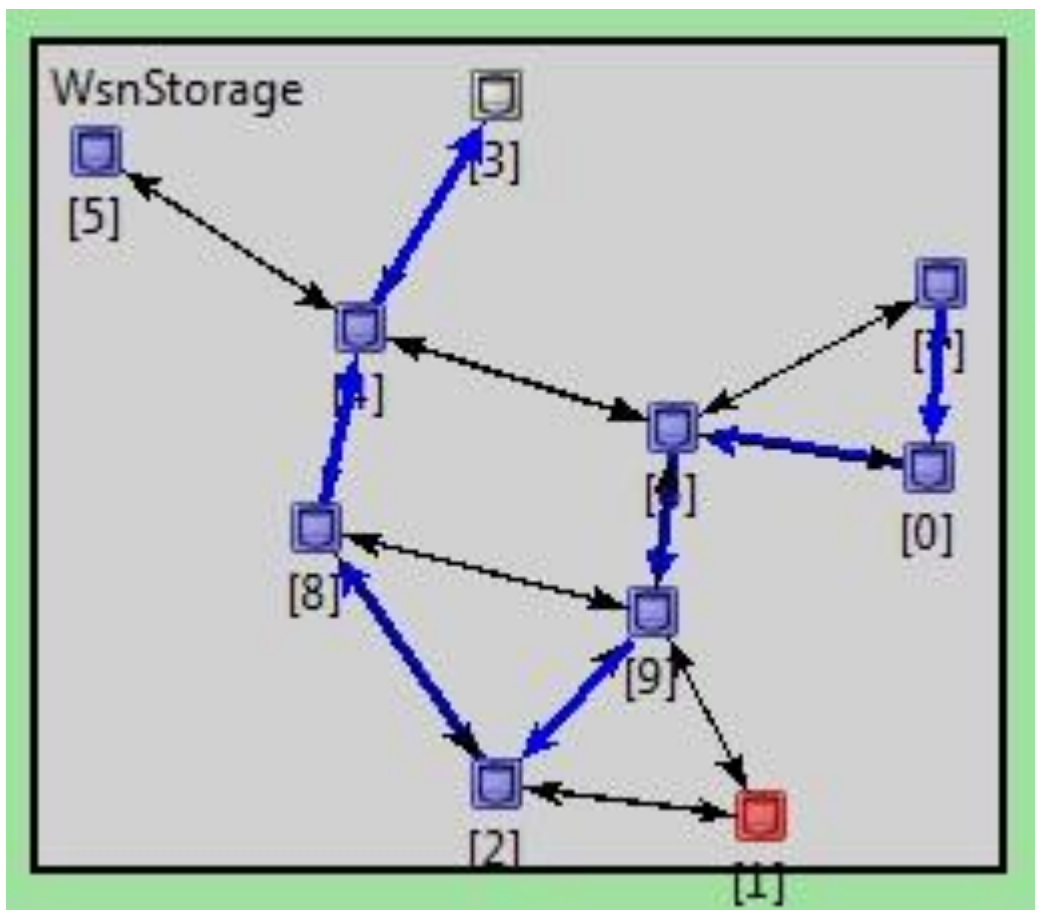

(b) Tree constructed in this paper

Figure 8 Construction of Repair Tree

In the two figures, red node refers to the failed node, blue nodes refer to the surviving nodes, and white node refers to the newcomer. Fig. 8(a) shows the tree required in the traditional method, and Fig. 8(b) shows the tree constructed in this paper. This experiment adopted $(n,7)$ MDS codes for implementation, so repair of one storage node required at least 7 storage nodes to participate in the repair process. In Fig. 8(b), a repair tree without any branch can be smoothly constructed. Such a tree may not exist in some cases due to the location of the newcomer, in which case, the construction algorithm will construct a suboptimum tree which will have some branches. Under this circumstance, the computation and transmission overhead of the node connected to the newcomer is slightly higher than other nodes, but this situation is inevitable sometimes.

Then we will discuss the energy efficiency of the repair technology. Similarly, we will continue evaluating it on the two aspects of total energy consumption and the load balancing of energy. The overhead generated by using the repair method based on network coding and that of the traditional repair method are as the following:

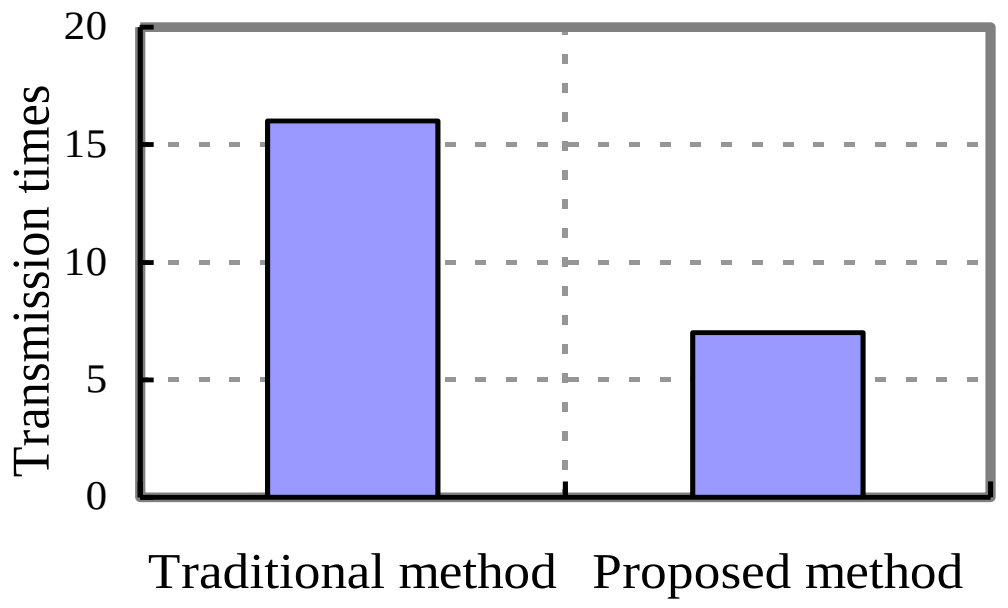

(a) Total transmission times

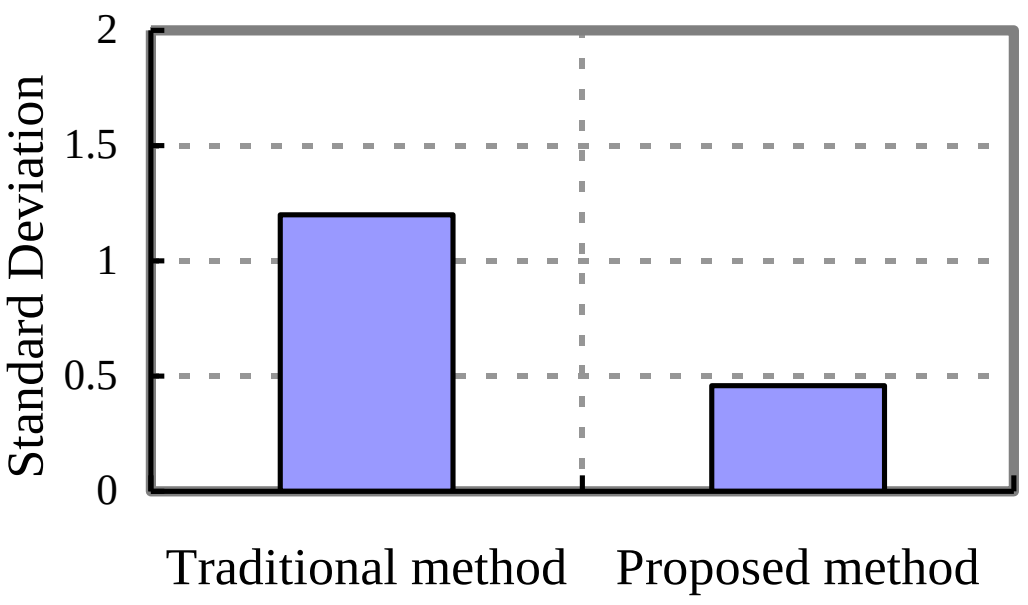

(b) Load balancing of energy

Figure 9 Comparison of Energy Efficiency

In accordance with the simulation results, we notice that after using the repair technology based on network coding, both the total energy consumption and load balancing have been improved. Both the two advantages are benefiting from the re-encoding at intermediate nodes. We tried to compare our repair method with previous methods such as MSR and MBR [19] which are also based on network coding. However, we failed to make the comparison since we observe that our method and the pervious methods are not comparable. Although both our method and previous method are based on network coding, they are completely different in topology. When MSR or MBR is employed, it is required to construct multiple paths from the newcomer to different storage nodes, and the connections among storage nodes are not required. When our method is employed, only one single path is required to connect to the newcomer, but the storage nodes are required to be able to communicate with each other. Therefore, the network topologies are completely different, and we consider that the comparison based on different topologies is unnecessary.

### 6.3 Computational Complexity

Objectively speaking, re-encoding operations at intermediate nodes will introduce additional energy consumption. In order to clarify the computation complexity of re-encoding operation in WSNs, we conducted some experiments to evaluate the computational complexity of network coding in practical applications. The following table shows the main properties of the sensor node we employed.

Table 1: Properties of the sensor node

| Parameter | Value | Parameter | Value |
|---|---|---|---|
| Chip | MC13213 | Standard | 802.15.4 |
| CPU | HCS08 | Memory | 4KB |
| Frequency | 40MHz | Flash | 16KB |

Based on this sensor node, we obtain some results.

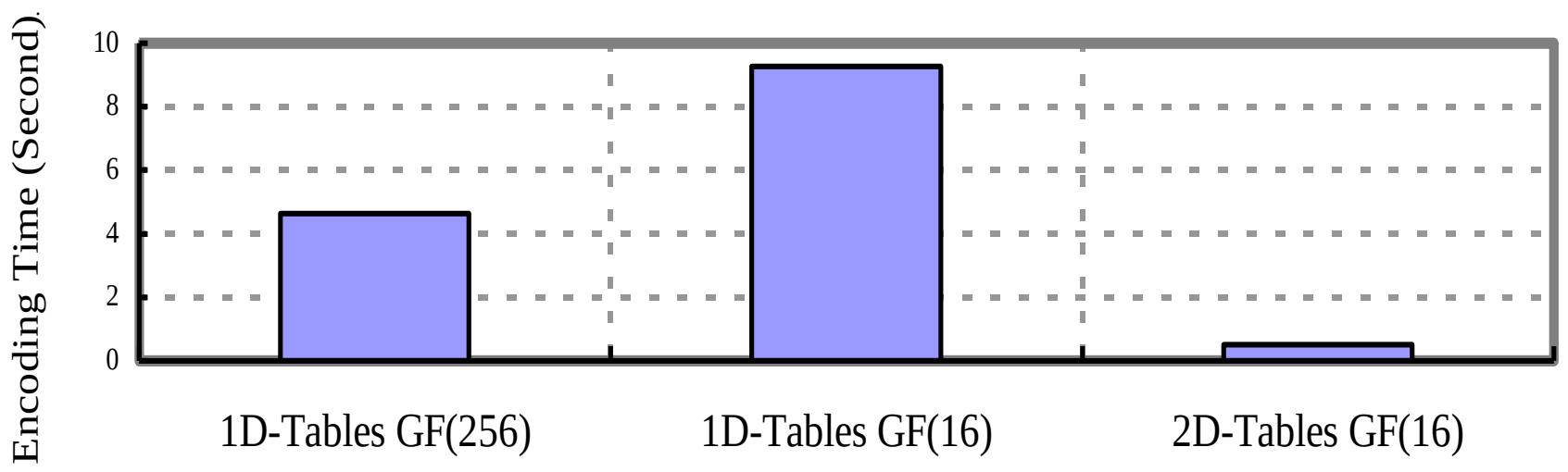

Figure 10 Computational Complexity of Network Coding

There are three experimental results, and 1M bits were encoded for each experiment. In these experiments, the network coding operations were based on look-up tables since the computation overhead of repeated calculation is unacceptable. Therefore, we made a multiplication table and a logarithm table to accelerate the computation speed of network coding, but there are some differences among these experiments. For the first experiment, the finite field $GF(2^8)$ was employed, and the tables we constructed are 1 dimensional. For the second experiment, the finite field $GF(2^4)$ was employed, and the two tables were 1 dimensional as well. We observed that the computation overhead over a small finite field was greater than that over a large field. For the third experiment, we also employed $GF(2^4)$ to be the finite field, but the difference is that the constructed tables were 2-dimensional. We observed that, after using the 2-dimensional look-up tables, the overhead of network coding was much smaller.

From the perspective of practical application, using 2-dimensianl tables or computing over a relatively large finite field will increase the computation efficiency since current processors are byte-oriented. However, these two strategies may not be employed at the same time since constructing 2-dimensional tables for a large finite field will cost too much storage space. If we construct 2-dimensional tables for the field $GF(2^8)$, it will require a memory of 128Kbytes to store the two tables. Note that the sensor node we employed only has 4 KB memory space, so it is impossible to construct the 2-dimensional tables over $GF(2^8)$. By the way, the storage spaces required in the three experiments are 512 bytes, 32 bytes and 512 bytes respectively.

Overall, we consider that a tradeoff between storage space overhead and computation overhead can be found, and then the overhead of network coding is affordable. As shown in Fig. 10, half a second is sufficient to encode 1M bit data, and we believe the encoding efficiency is acceptable for many applications in WSNs.

## 7 Conclusion

This paper has applied optimization encoding method at the source node and the re-encoding idea at intermediate nodes from the network coding theory into the distributed data storage and repair technologies in WSNs. Both of the proposed distributed data storage technology and repair technology can reduce the total energy consumption of the WSNs without reducing the system service level, and in the meantime, these technologies can balance the energy consumption of networks. In addition, the proposed scheme has also made improvement on the aspects of system buffer overhead and system security. Moreover, this paper shows that this scheme can work over a very small finite field, and the explicit lower bound of finite field required in this paper is also provided. Finally, some simulations were conducted to verify the proposed technologies, and some experiments were carried out to evaluate the computational complexity of network coding in practical scenarios.

## Acknowledgment

The work in this paper is supported by Key Technology R&D Program of Jiangsu, China (SBE201230225, SBE201000478, BE2011342). High School Research Industrialization Project of Jiangsu, China (JHZD2012-2) “333” Project of Jiangsu, China (AB41080)